# Pressure dependent physical properties of a potential high-$T_\text{C}$ superconductor ScYH$_6$: insights from first-principles study

Md. Ashraful Alam[1,2], F. Parvin[2], S. H. Naqib[2]*

[1]*Department of Physics, Mawlana Bhashani Science and Technology University, Santosh, Tangail 1902, Bangladesh*

[2]*Department of Physics, University of Rajshahi, Rajshahi 6205, Bangladesh*

*\*Corresponding author*: Email*: salehnaqib@yahoo.com*

**Abstract**

We have investigated the structural, elastic, electronic, thermophysical, superconducting, and optical properties of ScYH$_6$ under uniform hydrostatic pressures up to 25 GPa, using the density functional theory (DFT) formalism. Most of results reported here are novel. The compound ScYH$_6$ has been found to be elastically and thermodynamically stable within the pressure range considered. The compound is brittle; the brittleness decreases with increasing pressure. The elastic anisotropy is low and the machinability index is moderate which increases gradually with rising pressure. The compound is a hard material. The electronic band structure shows weakly metallic character with low density of states at the Fermi level. The Debye temperature of the compound is high and increases with increasing pressure. The Grüneisen parameter of ScYH$_6$ is low and the phonon thermal conductivity is high at room temperature. The compound is a very efficient reflector of infrared radiation. The compound is also an efficient absorber of visible and ultraviolet light. The overall effect of pressure on optical parameters is small. We have also investigated the pressure induced changes in the predicted superconducting state properties by considering the changes in the electronic density of states at the Fermi level, Debye temperature, and the repulsive Coulomb pseudopotential. The superconducting transition temperature is found to increase gradually with increasing pressure.

## 1. Introduction

Usually compounds in which one or more hydrogen atoms are present are known as hydrides. Metallization of pure hydrogen needs high pressure (up to 500 GPa). On the other hand hydrogen rich binary hydrides can be metalized under much lower pressures in comparison to that required for pure hydrogen [1]. Nowadays, metallic hydrides have drawn the attention of scientific community due to their high temperature superconducting properties such as in $H_3S$ [2,3] ($T_C$ ~203 K under 155 GPa), $YH_9$ with a $T_C$ of 253−276 K; stable at 200 GPa [4], $YH_6$ with a $T_C$ of 251−264 K at 110 GPa [5], $YH_{10}$ [6] with a $T_C$ ~326 K at 250 GPa. All these hydrides have high Debye temperature and strong electron-phonon coupling [7]. There are various other applications of hydrides such as in cryo-coolers [8], catalysis [9] and as chemical hydrogen storage material and in nuclear technology industry [10–14]. Transition metal hydrides focus on searching for potential superconductors with high $T_C$ under pressure [15-20]. For example, PdH [15] exhibits a moderately high superconducting transition temperature ($T_C$). On the other hand, there are a large number of ternary hydrides predicted to possess very good superconducting properties under pressure such as: $ScCaH_8$ and $ScCaH_{12}$ with the corresponding forecasted $T_C$ ~212 K and ~182 K, respectively, at 200 GPa [21], $ScYH_6$ with $T_C$ ~32.11 K to 52.90 K in the pressure range 0-200 GPa [1], $H_3SXe$ with a $T_C$ of 89 K at 240 GPa, $MgGeH_6$ with a $T_C$ ~67 K at 200 GPa [22], $CaYH_{12}$ with a $T_C$ ~258 K at 200 GPa [23], $LaSH_6$ with a superconducting transition temperature of ~35 K at 300 GPa [24], $MgSiH_6$ with a $T_C$ ~63 K at 250 GPa [25], $MgScH_6$ with a $T_C$ ~41 K at 100 GPa [26], and $MgVH_6$ with a $T_C$ of ~27.6 K at 150 GPa [27]. There are many other binary hydrides as well. Kong et al. observed that the $T_C$ of $YH_9$ is about 243 K at 201 GPa [16]. Previous investigations have established that introducing extra electrons via metal doping into known hydrogen-rich binary hydrides can effectively tune the metallization pressure and superconducting behavior of compounds, opening a new possibility in the quest for novel high-temperature superconductors under lower pressures [28]. Wei et al. [1] selected yttrium hydrides as the parent hydrides. In yttrium hydrides, scandium was doped to form new Sc-Y-H crystal systems under pressure [1]. Yttrium and scandium are transition metals. Having low mass of hydrogen, including the zero-point energy contributions which might be essential in determining the structural stability of hydrogen-rich compounds [29]. The calculations indicated that the doping of Sc in yttrium hydrides results in a decline of $T_C$; on the other hand, the synthesis of new Sc-Y-H ternary hydrides is much easier compared with binary yttrium hydrides. The findings revealed that ternary hydrides are promising candidates in the search of new high-$T_C$ superconductors, which can be synthesized under milder pressure conditions compared to many other hydrogen-rich metallic systems.

In this work, we have explored the elastic, hardness, thermal, and optoelectronics properties of $ScYH_6$ under pressure using the ab-initio methodology for the first time. Electronic band structure properties are revisited and pressure dependent superconducting state properties are also investigated.

## 2. Computational scheme

For geometry optimization of the compound of interest, plane wave pseudopotential [30] method was used which is contained in the CASTEP code [31]. For the electronic exchange-correlation functionals local density approximation (LDA) [32] was used. For the calculations of electron-ion interactions ultrasoft pseudopotential was used [33]. The BFGS algorithm was used to minimize the total energy and internal forces [34] within the optimized crystal structure. Monkhorst–Pack grid [35] was used for *k*-point sampling. The elastic constants were determined using the *stress-strain* module in the CASTEP. Thermophysical parameters were computed from the elastic constants and moduli. The tolerance levels and parameter settings used during computations are: *k*-mesh size of 15×15×15 for the Brillouin zone (BZ) sampling; plane wave cutoff energy of 500 eV is considered for the basis set. The total energy convergence tolerance is $5\times10^{-6}$ eV/atom, maximum force tolerance of 0.01 eV/Å, maximum stress of 0.02 GPa, maximum ionic displacement is $5 \times 10^{-4}$ Å and the self-consistence field tolerance is set to $5\times10^{-7}$ eV/atom for all of the pressures considered in our calculations. In these calculations the valence electrons for various atoms were as follows: Sc: $3s^2\ 3p^6\ 3d^1\ 4s^2$; Y: $4s^2\ 4p^6\ 4d^1\ 5s^2$; H: $1s^1$.

## 3. Results and analysis

3.1 *Structure and stability*

Figure 1 shows 3D crystal structure of $ScYH_6$. In the $ScYH_6$ structure, yttrium (Y) atom occupies 1a site with Wyckoff position (0.5, 0.5, 0.5) which is located at the center of the structure. Scandium (Sc) atoms occupy 1a site with Wyckoff position (0, 0, 0) which are located at the eight corners in the structure and the hydrogen (H) atoms occupy 6f site with Wyckoff position (0.25, 0, 0.5) which are located at six faces in the crystal structure. $ScYH_6$ (space group Pm-3) structure contains one formula unit i.e. there are 8 atoms in total in a unit cell. The optimized unit cell parameters (*a, b, c* and *V*), the cohesive energy/atom $E_{coh}$, and the enthalpy of formation ($\Delta H$) at different hydrostatic pressures are summarized in Table 1.

In this work, we have calculated the cohesive energy per atom ($E_{coh}$) using the approach adopted in Refs.[36–38] to determine the chemical stability. The $E_{coh}$ has been computed using the following equation:

$$E_{coh} = \frac{E_{ScYH_6} - E_{Sc} - E_Y - 6E_H}{8} \qquad (1)$$

where, $E_{ScYH_6}$ is total energy per formula unit of ScYH$_6$ and $E_{Sc}$, $E_Y$ and $E_H$ are the total energies of isolated single Sc, Y and H atom, respectively. From Table 1, it is found that the values of cohesive energy per atom and formation enthalpy are negative, which suggest that the structures of ScYH$_6$ is thermodynamically stable [39]. Moreover, our structural parameters are in good agreement with the previous results [1].

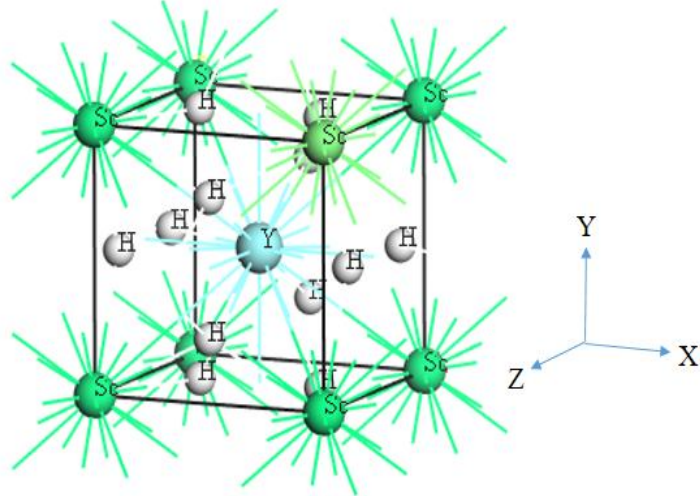

**Figure 1**: Three dimensional (3D) schematic crystal structure of ScYH$_6$.

**Table 1**. Structural parameters (*a, b, c* and *V*), cohesive energy/atom ($E_{coh}$), and formation enthalpy (Δ*H*) of ScYH$_6$ under different pressures.

| Compound | Pressure, *P* (GPa) | Lattice parameters (Å) | | | Volume, *V* (Å³) | Cohesive energy, $E_{coh}$ (eV/atom) | Formation enthalpy, Δ*H* (eV/atom) |
|---|---|---|---|---|---|---|---|
| | | *a* | *b* | *c* | | | |
| ScYH$_6$ | 0 | 3.9039 | 3.9039 | 3.9039 | 59.50 | -4.51 | -110.23 |
| | 5 | 3.8486 | 3.8486 | 3.8486 | 57.00 | -4.50 | -110.22 |
| | 10 | 3.8010 | 3.8010 | 3.8010 | 54.92 | -4.49 | -110.21 |
| | 15 | 3.7593 | 3.7593 | 3.7593 | 53.13 | -4.47 | -110.19 |
| | 20 | 3.7219 | 3.7219 | 3.7219 | 51.56 | -4.45 | -110.17 |
| | 25 | 3.6882 | 3.6882 | 3.6882 | 50.17 | -4.43 | -110.15 |

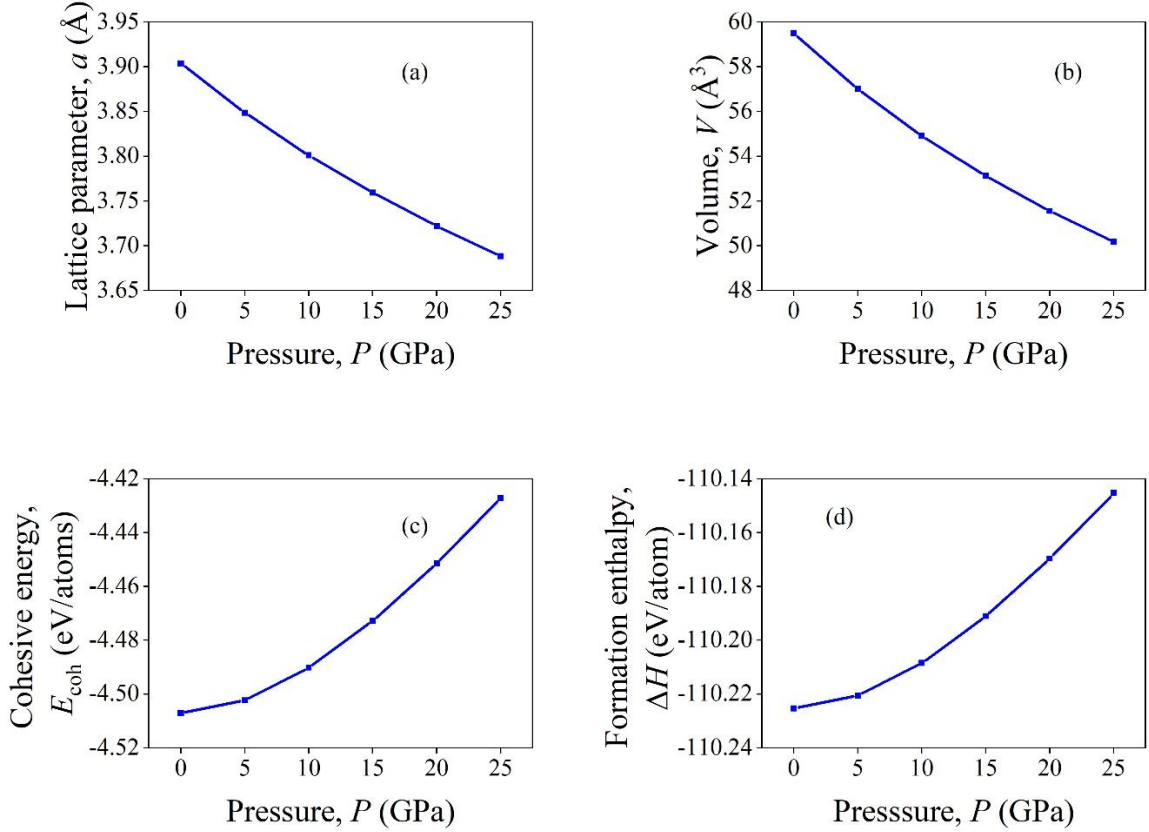

**Figure 2**: Lattice parameters (a) and (b), cohesive energy (c) and enthalpy (d) of ScYH$_6$ under pressure.

From Table 1 (cf. Fig.2) it is observed that lattice parameters ($a = b = c$ and $V$) decrease in same pattern with pressure, and cohesive energy decreases with pressure also. We don't see any rapid change in lattice parameters at any pressure. So, it is expected that there is no structural phase transition within the pressures considered in this work.

*3.2 Elastic properties*

*3.2.1 Single crystal elastic constants*

For a cubic crystal, there are three independent elastic constants: $C_{11}$, $C_{12}$, and $C_{44}$. All of these elastic constants are recorded in Table 2. From Table 2 [cf. Fig.3 (a)] it is observed that elastic constant $C_{ij}$ increases almost linearly up to 25 GPa. All the elastic constants $C_{ij}$ are positive. The values of $C_{11}$, $C_{22}$, and $C_{33}$ represent the [100], [010] and [001] directional resistance against uniaxial stress. Cubic symmetry demands $C_{11} = C_{22} = C_{33}$, which indicates incompressibility is the same along the three principal axes. Under hydrostatic pressure, the stability conditions of a cubic crystal are as follows [40]:

$$M_1 = \frac{(C_{11} + 2C_{12})}{3} + \frac{P}{3} > 0, \ M_2 = (C_{44} - P) > 0, \ M_3 = [\frac{(C_{11} - C_{12})}{2} - P] > 0$$

From Table 2 it is observed that all the requirements of stability criteria under pressure are satisfied. The facts that $C_{11} > C_{12}$ and $C_{11} > C_{44}$ indicates axial bonding between nearest atoms is stronger than the bondings between atoms in different shear planes. The Cauchy pressure (CP) defined as ($C_{12}$ - $C_{44}$) can determine the angular character of atomic bondings in materials [41]. If the CP > 0, it indicates ductile nature and CP < 0 indicates brittle nature. From Table 2 it is observed that ScYH$_6$ is brittle up to 20 GPa and at 25 GPa it exhibits borderline ductile nature. The tetragonal shear modulus (TSM) [given by, 0.5($C_{11}$ - $C_{12}$)] is another useful elastic parameter. The value of TSM is used to measure the shear stiffness of a crystal i.e., the resistance to shear stress or shear deformation. TSM is also linked with the sound velocity in the solid. Positive values of TSM are suggestive of dynamical stability of the structure in the long wavelength limit. The positive values of TSM indicate dynamical stability of ScYH$_6$.

**Table 2**. Single crystal elastic constants ($C_{ij}$), Cauchy pressure ($C'$) and tetragonal shear modulus ($C''$) of ScYH$_6$ under different pressures.

| Compound | Pressure, $P$ (GPa) | $C_{ii}$ (GPa) | | | CP (GPa) | TSM (GPa) |
|---|---|---|---|---|---|---|
| | | $C_{11}$ | $C_{12}$ | $C_{44}$ | $C'$ | $C''$ |
| ScYH$_6$ | 0 | 224.16 | 50.35 | 87.02 | -36.67 | 86.90 |
| | 5 | 250.68 | 63.12 | 93.17 | -30.05 | 93.78 |
| | 10 | 276.92 | 75.33 | 97.71 | -22.38 | 100.79 |
| | 15 | 303.14 | 87.10 | 101.27 | -14.16 | 108.02 |
| | 20 | 329.68 | 98.24 | 103.62 | -5.38 | 115.72 |
| | 25 | 357.77 | 108.10 | 104.32 | 3.78 | 124.84 |

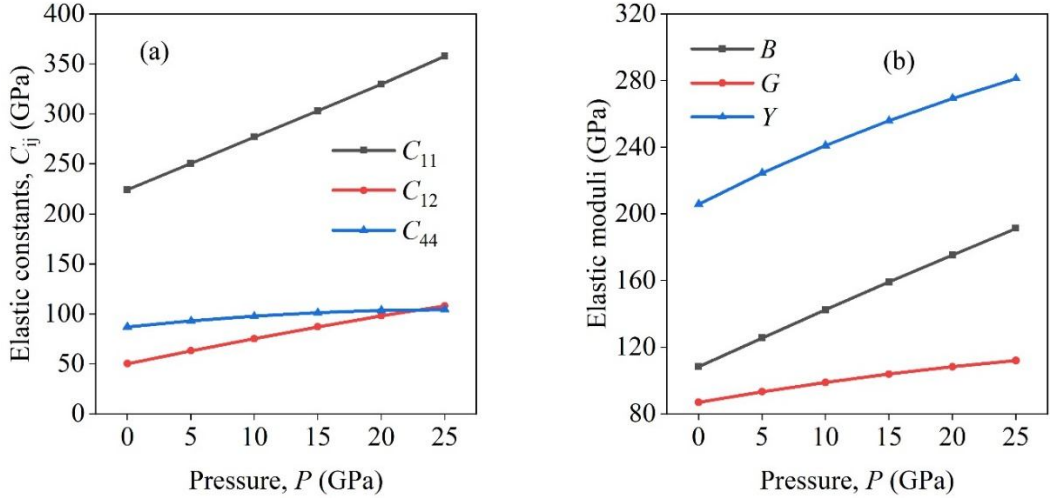

**Figure 3**: (a) Single crystal elastic constant and (b) Elastic moduli under pressure.

*3.2.2 Polycrystalline elastic properties*

The values of polycrystalline elastic moduli ($B$, $G$, and $Y$) which quantify the bulk elastic response of a solid are displayed in Table 3 [cf. Fig.3 (b)]. The symbols $B$, $G$, and $Y$ indicate bulk, shear, and Young modulus, respectively. The symbols $V$, $R$, and $H$ represent Voigt [42], Reuss [43] and Hill [44] approximations. The Hill approximated elastic moduli are the average of the Voigt and Reuss approximated elastic moduli. From Table 3 it is observed that elastic moduli increase with increasing pressure. The bulk modulus defines the resistance against volume change. The shear modulus defines the elastic resistance against shape changing stress. The Young modulus gives the measure of the resistance against the length changing stress. Among the three moduli, the shear modulus is the lowest at all pressures. This suggests that the elastic failure of $ScYH_6$ should be controlled by the shape changing strain.

**Table 3**. Polycrystalline elastic moduli of $ScYH_6$ under different pressures.

| Compound | Pressure, $P$ (GPa) | Polycrystalline elastic moduli (GPa) | | | | | | | | |
|---|---|---|---|---|---|---|---|---|---|---|
| | | Bulk, $B$ | | | Shear, $G$ | | | Young, $Y$ | | |
| | | $B_V$ | $B_R$ | $B_H$ | $G_V$ | $G_R$ | $G_H$ | $Y_V$ | $Y_R$ | $Y_H$ |
| ScYH$_6$ | 0 | 108.29 | 108.29 | 108.29 | 86.97 | 86.97 | 86.97 | 205.81 | 205.81 | 205.81 |
| | 5 | 125.64 | 125.64 | 125.64 | 93.41 | 93.41 | 93.41 | 224.58 | 224.58 | 224.58 |
| | 10 | 142.53 | 142.53 | 142.53 | 98.94 | 98.92 | 98.93 | 241.05 | 241.01 | 241.03 |
| | 15 | 159.12 | 159.12 | 159.12 | 103.97 | 103.86 | 103.92 | 256.12 | 255.91 | 256.02 |
| | 20 | 175.38 | 175.38 | 175.38 | 108.46 | 108.14 | 108.30 | 269.77 | 269.12 | 269.44 |
| | 25 | 191.32 | 191.32 | 191.32 | 112.52 | 111.66 | 112.09 | 282.24 | 280.42 | 281.33 |

**Table 4**. Calculated Poisson's ratio ($v$), Pugh's ratio ($G/B$) and machinability index ($\mu_m$) of ScYH$_6$.

| Compound | Pressure, $P$ (GPa) | $v$ | $G/B$ | $\mu_m$ |
|---|---|---|---|---|
| ScYH$_6$ | 0  | 0.18 | 0.80 | 1.24 |
|          | 5  | 0.20 | 0.74 | 1.35 |
|          | 10 | 0.22 | 0.69 | 1.46 |
|          | 15 | 0.23 | 0.65 | 1.57 |
|          | 20 | 0.24 | 0.62 | 1.69 |
|          | 25 | 0.25 | 0.59 | 1.83 |

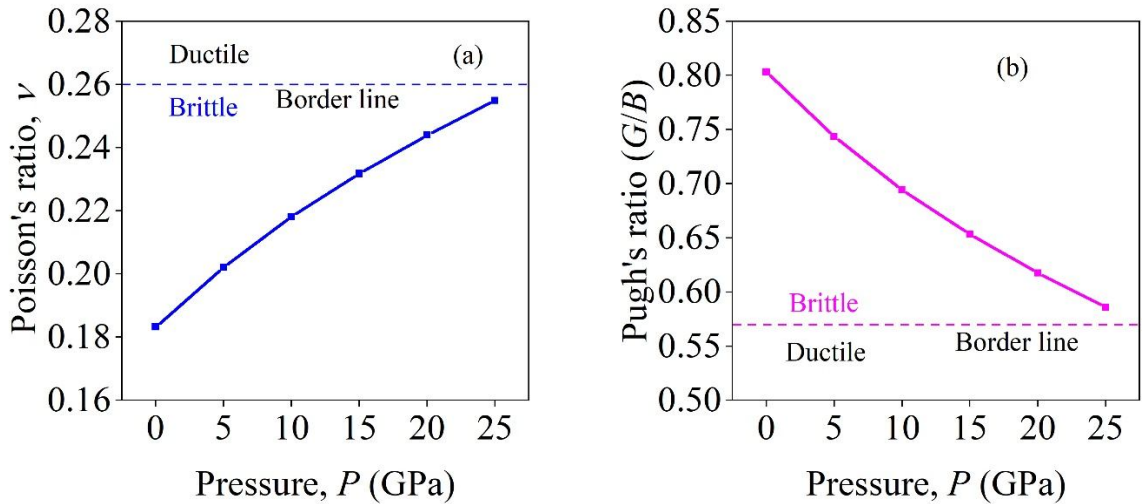

**Figure 4**: (a) Poisson's ratio and (b) Pugh's ratio of ScYH$_6$ under pressure.

The Poisson's [45] and Pugh's ratio [46] define the brittle and ductile natures of a solid. According to Frantsevich [45], if the Poisson's ratio $v > 0.26$, the solid exhibits ductile nature and on the other hand, if $v < 0.26$ it exhibits brittle nature. From Table 4 it is observed that ScYH$_6$ exhibits brittle in nature. The level of brittleness decreases with increasing pressure. For the Pugh's ratio, if $G/B > 0.57$, the solid should exhibit brittle nature otherwise, it should be ductile. From Table 4 [cf. Fig. 4 (b)] it is observed that Pugh's ratio gives the same information as Poisson's ratio about the brittleness of ScYH$_6$. Both these ratios approach borderline ductile value at 25 GPa. The machinability index ($\mu_m = B/C_{44}$) is used as an indicator of dry lubricity of solids. High machinability index indicates that the solid is easy to shape using cutting tools and have low level of frictional loss. The machinability index increases systematically with increasing pressure. From Table 4 it is observed that at high pressures $ScYH_6$ should have fair level of dry lubricity.

*3.3 Elastic Anisotropy*

**Table 5**. The shear anisotropy factors $A$, and $A^B$ (in %), $A^G$ (in %), $A^U$ of $ScYH_6$.

| Compound | Pressure, $P$ (GPa) | $A$ | $A^B$ | $A^G$ | $A^U$ |
|---|---|---|---|---|---|
| ScYH$_6$ | 0 | 1.00 | 0.00 | 0.00 | 0.00 |
|  | 5 | 0.99 | 0.00 | 0.00 | 0.00 |
|  | 10 | 0.97 | 0.00 | 0.01 | 0.00 |
|  | 15 | 0.94 | 0.00 | 0.05 | 0.00 |
|  | 20 | 0.90 | 0.00 | 0.15 | 0.01 |
|  | 25 | 0.84 | 0.00 | 0.39 | 0.04 |

Elastic anisotropy is important to know to understand the direction dependent bonding characteristics and mechanical properties of crystalline solids. Shear elastic anisotropy factors are one for an isotropic crystal, while any value except unity is a measure of the degree of anisotropy possessed by the crystal. The single crystal shear anisotropy factors $A_1$, $A_2$, and $A_3$ are computed for cubic symmetry using the following equations [47-49]:

$$A_1 = \frac{4C_{44}}{C_{11} + C_{33} - 2C_{13}} = A_2 = \frac{4C_{55}}{C_{22} + C_{33} - 2C_{23}} = A_3 = \frac{4C_{66}}{C_{11} + C_{22} - 2C_{12}} = A \qquad (2)$$

All the computed values of $A_1$, $A_2$, and $A_3$ are unity in the ground state. This is a consequence of cubic symmetry of $ScYH_6$. The indices $A^B$ and $A^G$ are the percentage anisotropies in compressibility and shear, respectively, and $A^U$ is the universal anisotropy index. The zero values of $A^B$, $A^G$ and $A^U$ represent elastic isotropy of a crystal and their non-zero values represent anisotropy [50,51]. The indices $A^B$, $A^G$ and $A^U$ are calculated using the following equations:

$$A^B = \frac{B_V - B_R}{B_V + B_R}, \quad A^G = \frac{G_V - G_R}{G_V + G_R}, \text{ and } A^U = 5\frac{G_V}{G_R} + \frac{B_V}{B_R} - 6 \geq 0 \qquad (3)$$

The computed values of $A$, $A^U$, $A^B$, and $A^G$ are listed in Table 5. The values in Table 5 show clearly that the $ScYH_6$ is elastically very close to isotropic at low pressures, but at high pressures it is slightly anisotropic. Irrespective of pressure, the bulk modulus/compressibility remains completely isotropic. Anisotropies in elastic behavior are clear indications that atomic bonding strengths in different directions in the crystals are different.

*3.4 Sound velocities and hardness*

Large number of thermophysical parameters including the Debye temperature of a crystal is closely related to sound velocities [52].

**Table 6**. Calculated density ($\rho$), transverse sound velocities ($v_t$), longitudinal sound velocities ($v_l$) and average sound velocities ($v_m$) of ScYH$_6$.

| Compound | Pressure, $P$ (GPa) | Density, $\rho$ (gm/cm$^3$) | Sound Velocities (km/s) | | |
|---|---|---|---|---|---|
| | | | $v_t$ | $v_l$ | $v_m$ |
| ScYH$_6$ | 0 | 3.905 | 4.7193 | 7.5781 | 5.2008 |
| | 5 | 4.076 | 4.7874 | 7.8349 | 5.2864 |
| | 10 | 4.231 | 4.8358 | 8.0542 | 5.3492 |
| | 15 | 4.373 | 4.8759 | 8.2513 | 5.4018 |
| | 20 | 4.506 | 4.9024 | 8.4241 | 5.4388 |
| | 25 | 4.631 | 4.9199 | 8.5784 | 5.4652 |

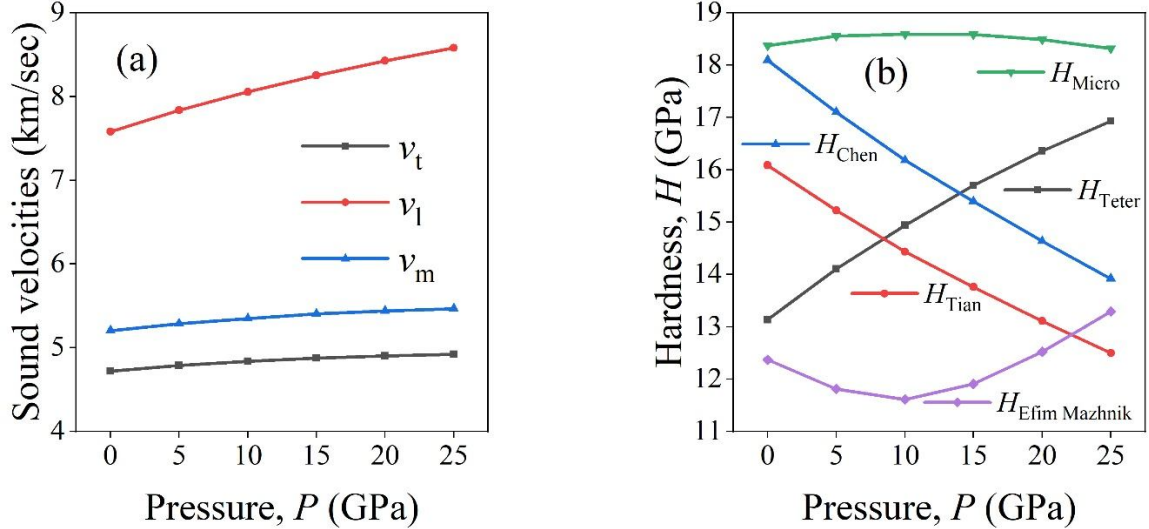

**Figure 5**: (a) Sound velocities and (b) Hardness of ScYH$_6$ under pressure.

Crystalline solids support both longitudinal and transverse modes of propagation of acoustic disturbances. The phonon thermal conductivity of solids and sound velocities vary in the same mode. The calculated values of sound velocities ($v_t, v_l$ and $v_m$) under pressure of ScYH$_6$ are gathered in Table 6 [cf. Fig.5 (a)]. The sound velocities were determined using the following equations [53-56]:

$$v_t = \sqrt{\frac{G}{\rho}}, \quad v_l = \sqrt{\frac{3B + 4G}{3\rho}} \text{ and } v_m = \left[\frac{1}{3}\left(\frac{2}{v_t^3} + \frac{1}{v_l^3}\right)\right]^{-\frac{1}{3}} \quad (4)$$

Where, $v_t, v_l$ and $v_m$ signify the transverse, longitudinal, and the mean sound velocities, respectively. In general, the sound velocities increase with increasing pressure due to the increase in the crystal stiffness. The longitudinal sound velocities are larger than that of transverse velocity for each direction of propagation. This is due to the fact that $C_{11}$ is greater than $C_{12}$ or $C_{44}$ for $ScYH_6$.

Hardness is used to identify elastic and plastic behavior of a solid. In our study, we have calculated the hardness of cubic $ScYH_6$ at different pressures using the formalisms developed by Teter et al. [57], Tian et al. [58], Chen et al. [59], the microhardness [60], and that by Efim Mazhnik [61]. The calculated values of hardness are tabulated in Table 7. Hardness values vary in different ways with pressure in different formalisms. The microhardness is almost pressure insensitive with reasonably high value. We can say that $ScYH_6$ is a hard compound.

**Table 7**. The calculated hardness of $ScYH_6$ at different pressures.

| Compound | Pressure, $P$ (GPa) | Hardness $H$ (GPa) | | | | |
|---|---|---|---|---|---|---|
| | | $H_{Teter}$ | $H_{Tian}$ | $H_{Chen}$ | $H_{micro}$ | $H_{Efim\ Mazhnik}$ |
| $ScYH_6$ | 0 | 13.13 | 16.09 | 18.10 | 18.37 | 12.37 |
| | 5 | 14.11 | 15.22 | 17.10 | 18.55 | 11.81 |
| | 10 | 14.94 | 14.43 | 16.18 | 18.59 | 11.61 |
| | 15 | 15.70 | 13.76 | 15.39 | 18.59 | 11.91 |
| | 20 | 16.35 | 13.11 | 14.63 | 18.49 | 12.52 |
| | 25 | 16.93 | 12.50 | 13.92 | 18.31 | 13.29 |

*3.5 Thermophysical properties*

A number of technologically important thermophysical properties of $ScYH_6$ are studied in this section.

*Grüneisen parameter ($\gamma$)*: The Grüneisen parameter $\gamma$ is an important lattice dynamical parameter which is closely related to thermophysical quantities that link the vibrational properties with the structural ones. It is related also to the thermal expansion coefficient, bulk modulus, specific heat, and electron-phonon coupling in solids. The normal thermal expansion of solids due to anharmonicity of interatomic forces is understood from the Grüneisen constant as well. There is a very important relation between the Grüneisen parameter and the Poisson ratio $v$ (the lateral strain coefficient) that characterizes the tendency of material towards retaining its initial volume in the course of elastic deformation and is defined by the relation between Grüneisen parameter and Poisson's ratio as follows: $\gamma = \frac{3}{2}\frac{1+v}{2-3v}$ [62]. Lower limit of Poisson's ratio i.e., $v = -1.0$ corresponds to a completely harmonic solid. High value of Grüneisen parameter is sometimes indicative of strong electron-phonon interaction in superconductors [63]. The pressure dependent values of the Grüneisen parameter of ScYH$_6$, calculated using the Poisson's ratio, are tabulated in Table 8. The obtained values of $\gamma$ are typical [64]. The values of the Grüneisen parameter are increasing with increasing pressure (Table 8).

**Table 8**. Calculated Grüneisen parameter ($\gamma$), Kleinmen parameter ($\zeta$), Debye temperature $\theta_D$ (K) and melting temperature $T_m$ (K) of ScYH$_6$ at different pressures.

| Compound | Pressure, $P$ (GPa) | $\gamma$ | $\zeta$ | $\theta_D$ | $T_m$ |
|---|---|---|---|---|---|
| ScYH$_6$ | 0 | 1.22 | 0.38 | 793.31 | 1362.71 |
| | 5 | 1.29 | 0.40 | 817.95 | 1482.05 |
| | 10 | 1.36 | 0.42 | 838.02 | 1600.16 |
| | 15 | 1.42 | 0.44 | 855.66 | 1718.13 |
| | 20 | 1.47 | 0.45 | 870.17 | 1837.56 |
| | 25 | 1.52 | 0.45 | 882.38 | 1963.97 |

*Kleinmen parameter ($\zeta$)*: This parameter quantifies the nature of internal strain and was introduced by Kleinman [65] to describe the relative ease of bond bending versus the bond stretching. Low level of bond bending contribution leads to $\zeta = 0$; while low level of bond stretching leads to $\zeta = 1$. The Kleinman parameter is calculated from the elastic constants $C_{11}$ and $C_{12}$ as follows:

$$\zeta = \frac{C_{11} + 8C_{12}}{7C_{11} + 2C_{12}} \tag{5}$$

The pressure dependent values of $\zeta$ are given in Table 8. It is clear from this table that both bond bending and bond stretching contributes to the overall mechanical strength of $ScYH_6$.

*Debye temperature ($\theta_D$)*: The Debye temperature $\theta_D$ is closely related to many physical properties of solids such as specific heat, melting temperature, thermal conductivity, hardness of solids, elastic constants, acoustic velocity, resistivity etc. Debye temperature provides information about the electron-phonon coupling and Cooper pairing mechanism of superconductivity. At low temperatures, the lattice vibrational excitations arise solely from acoustic modes. Thus the Debye temperature calculated by elastic constants is considered to be similar to that acquired from the specific heat measurements.

Debye temperature is calculated from the Anderson method [66] as follows:

$$\theta_D = \frac{h}{k_B}\left[\left(\frac{3n}{4\pi}\right)\frac{N_A \rho}{M}\right]^{1/3} v_m \tag{6}$$

where, $h$ is Planck's constant, $k_B$ is the Boltzmann's constant, $\rho$ is the density, $N_A$ is the Avogadro number, $M$ is the molecular mass, and $v_m$ is the average sound velocity. The mean sound velocity $v_m$ can be determined from the Eqn. 4:

From Table 8, it can be inferred that the $\theta_D$ increases monotonously with pressure. Usually, the increase of $\theta_D$ with pressure indicates the crystal stiffening, in the opposite case the system is driven effectively towards lattice softening. The Debye temperature of $ScYH_6$ is quite high, consistent with the hardness values obtained before.

*Melting temperature ($T_m$)*: The melting temperature is mostly used to gauge the overall bonding strength and limit of high temperature applicability of solids. The melting temperature of cubic compound is calculated with an empirical formula based on the single crystal elastic constants as follows [67]:

$$T_m = 354 + 1.5(2C_{11} + C_{33}) \tag{7}$$

The calculated $T_m$ is listed in Table 8. It is apparent that the melting point increases almost linearly with the increase of pressure. In general, compounds with high melting temperature have lower thermal expansion and high bonding energy and also have high Debye temperature. Melting temperature increases sharply with increasing temperature.

**Table 9**. Thermal expansion coefficient $\alpha$ ($10^{-5}$ K$^{-1}$), minimum thermal conductivity, $k_{min}$ (Wm$^{-1}$K$^{-1}$), lattice thermal conductivity, $k_{ph}$ (Wm$^{-1}$K$^{-1}$) at 300 K and fracture toughness, $K_{IC}$ (MPam$^{1/2}$) of ScYH$_6$.

| Compound | Pressure, $P$ (GPa) | $\alpha$ | $k_{min}$ | | $k_{ph}$ | $K_{IC}$ | |
|---|---|---|---|---|---|---|---|
| | | | Cahill | Clark | | Mazhnik | Alexander |
| ScYH$_6$ | 0 | 1.84 | 2.49 | 1.88 | 31.45 | 0.99 | 1.36 |
| | 5 | 1.71 | 2.62 | 1.97 | 30.13 | 1.16 | 1.50 |
| | 10 | 1.62 | 2.73 | 2.04 | 28.80 | 1.34 | 1.64 |
| | 15 | 1.54 | 2.84 | 2.11 | 27.68 | 1.52 | 1.76 |
| | 20 | 1.48 | 2.93 | 2.17 | 26.51 | 1.71 | 1.88 |
| | 25 | 1.43 | 3.02 | 2.22 | 25.37 | 1.89 | 1.99 |

*Thermal expansion coefficient* (TEC): Thermal expansion coefficient of a material is inter connected to many other temperature dependent physical properties, such as thermal conductivity, heat capacity, temperature variation of the energy band gap and electron effective mass. The TEC is denoted by $\alpha$. In this work the TEC of the material was calculated using the following equation [68]:

$$\alpha = \frac{1.6 \times 10^{-3}}{G} \tag{8}$$

The relation between thermal expansion coefficient and melting temperature is $\alpha \approx 0.02/T_m$ [68] i.e., TEC is inversely proportional to melting temperature. It is clear that high melting temperature has low thermal conductivity. Our calculated values are in good agreement with this statement. The TEC of ScYH$_6$ decreases steadily with increasing pressure (Table 9). This is a consequence of pressure induced stiffening of the crystal.

*Minimum thermal conductivity*: At temperatures above the Debye temperature, thermal conductivity of solids attains a minimum saturating value, known as the minimum thermal conductivity. It is denoted by $k_{min}$. According to the Cahill and Clarke model, minimum thermal conductivity can be calculated using the following equation [69]:

$$k_{min} = \frac{k_B}{2.48} n^{\frac{2}{3}}(v_l + v_{t1} + v_{t2}) \tag{9}$$

From this expression it is clear that minimum thermal conductivity depends on different sound velocities in different crystallographic directions.

For $k_{min}$, Clark also deduced the following equation [70]:

$$k_{min} = k_B v_m (V_{atomic})^{-\frac{2}{3}} \tag{10}$$

where, $k_B$ is the Boltzmann constant, $v_m$ is the average sound velocities and $V_{atomic}$ is the cell volume per atom. The minimum thermal conductivities of ScYH$_6$ compound are large [71,72]. We have also calculated the lattice thermal conductivity at 300 K following earlier studies [71,72] and presented those values in Table 9 together with minimum thermal conductivity.

*Fracture toughness* ($K_{IC}$): For functioning of a solid under mechanical strain and pressure, the fracture toughness as well as the hardness plays important roles. It determines the mechanical stability resistance of a solid to prevent the propagation of a crack produced inside. Quantitatively, it can be determined from the stress intensity factor at which a thin crack in the material begins to grow. According to Efim Mazhnik et al. [73] fracture toughness can be calculated using the following equations:

$$K_{IC} = a_0^{-0.5} V_0^{\frac{1}{6}} \xi(v) Y^{\frac{3}{2}} \tag{11}$$

Where, $\alpha_0 = 8840$, $V_0$ is the volume per atom. $\xi(v)$ is a dimensionless function of the Poisson's ratio defined as follows:

$$\xi(v) = \frac{1 - 13.7v + 48.6v^2}{1 - 15.2v + 70.2v^2 - 81.5v^3} \tag{12}$$

The computed fracture toughness at different pressures is given in Table 9. Alexander G. Kvashnin et al. [74] used the following equation for calculating fracture toughness:

$$K_{IC} = \alpha V_0^{\frac{1}{6}} G \left(\frac{B}{G}\right)^{\frac{1}{2}} \tag{13}$$

Where $\alpha = 1$ for hydride materials, $V_0$ is the volume per atom, $G$ and $B$ are the share and bulk moduli. The Alexander formalism results in higher values of fracture toughness of solids. The fracture toughness is positively correlated with pressure. $ScYH_6$ possesses high level of fracture toughness.

*3.6 Electronic band structure*

The calculated electronic band structures at equilibrium lattice constants under different pressures are illustrated along high symmetry directions in the first Brillouin zone in Figs. 6. From Fig. 6 it is observed that there is no band gap – implying the metallic character of $ScYH_6$. It is also observed that four distinct bands cross the Fermi level (placed at zero energy). Following Ref. [1], we have shown the electronic band structure of $ScYH_6$ at 0.01 GPa as well. At all the pressures considered, four bands cross the Fermi level. Among these four bands, three (at the $\Gamma$-point) are hole-like and the one at the M-point is electronic in nature. The band crossings are weak but they become more prominent at high pressures. Overall, we can characterize $ScYH_6$ as a metal with small Fermi surface. The degree of electronic dispersion is direction dependent within the Brillouin zone. From Fig. 6 it is seen that the bands crossing the Fermi level become more dispersive as pressure increases. Thus, the charge carrier effective mass should decrease gradually with increasing pressure.

The total and partial densities of states (TDOS and PDOS, respectively) are shown in Fig. 7. The values of TDOS at the Fermi level are 0.440, 0.440, 0.430, 0.181, 0.177, 0.174 and 0.180 states/eV-formula unit at 0 GPa, 0.01 GPa, 5 GPa, 10 GPa, 15 GPa, 20 GPa and 25 GPa, respectively. The TDOS at Fermi energy decreases significantly with pressure up to 20 GPa. The TDOS close to the Fermi level originated mainly from the H-*s*, Sc-*p* and Y-*p* electronic states. There is large hybridization among the electronic orbitals of Sc, Y, and H atoms close to the Fermi level. This suggests that covalent bondings should dominate in this compound. The Fermi level is located close to the center of a pseudogap. This suggests that the electronic stability of $ScYH_6$ is high. Compared to many other synthesized and prospective hydride superconductors, the TDOS at Fermi level of $ScYH_6$ is significantly smaller [75,76].

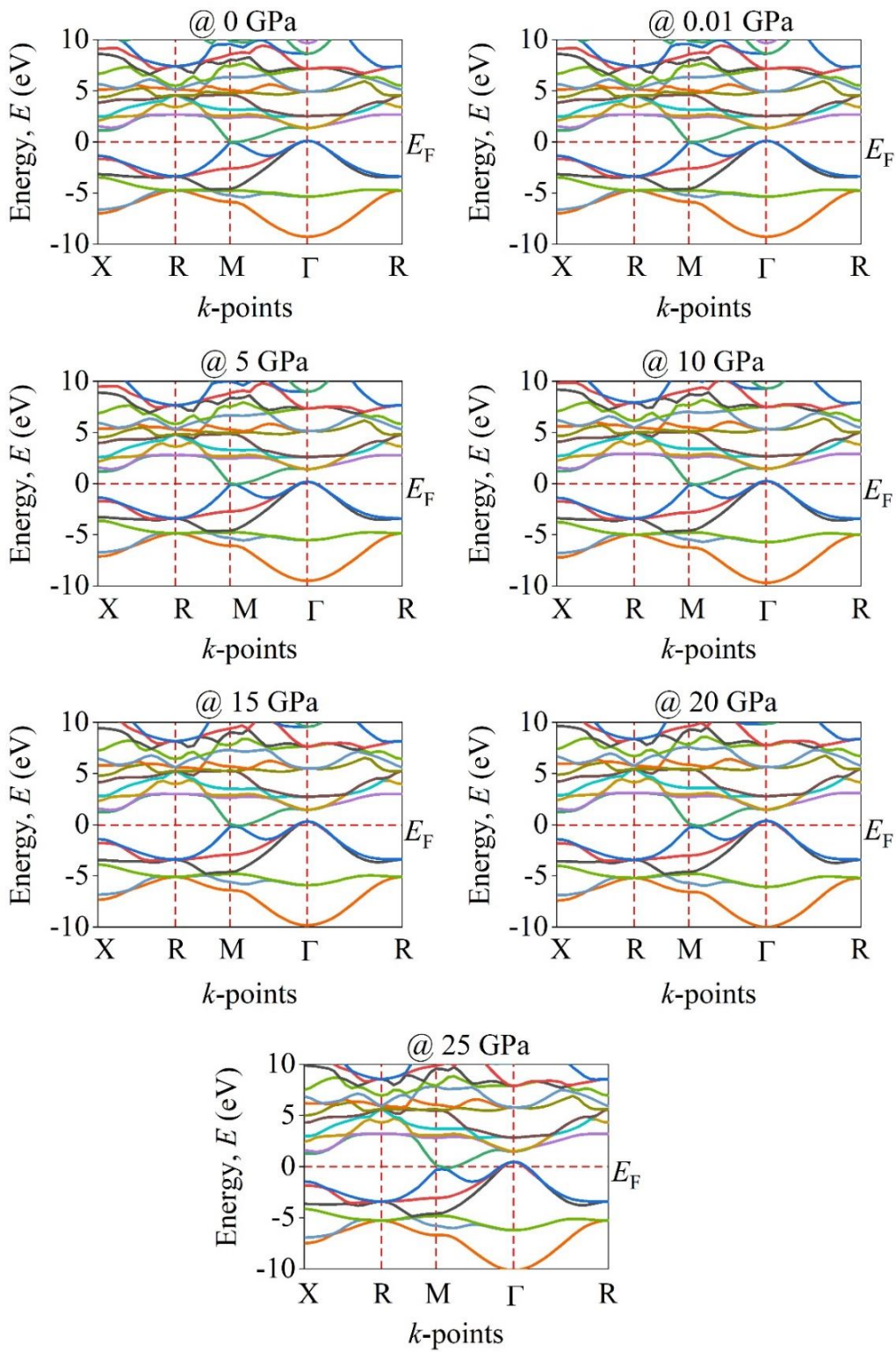

**Figure 6**: Electronic band structure of ScYH$_6$ under different pressures.

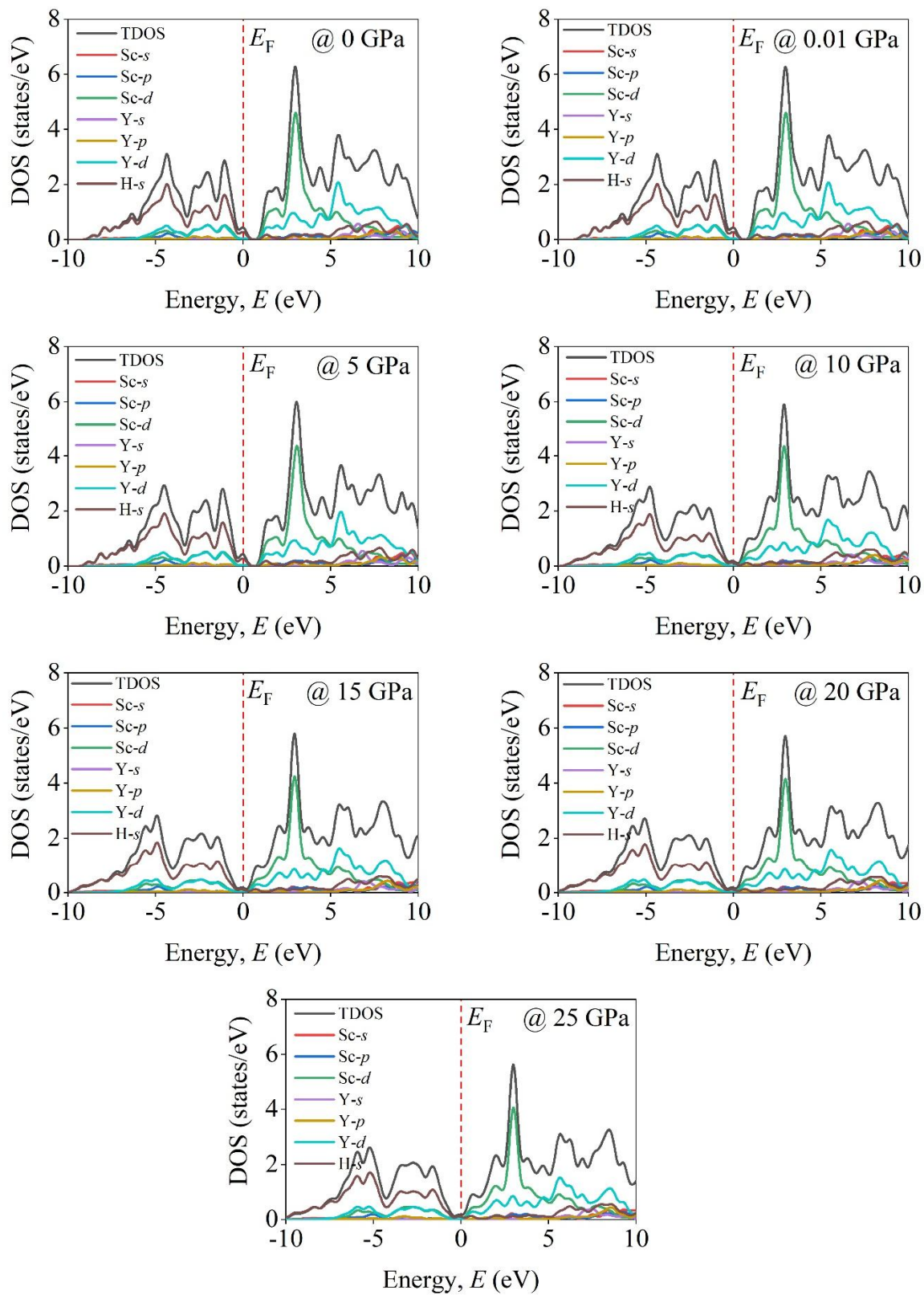

**Figure 7**: Density of states (TDOS and PDOS) of cubic ScYH$_6$ under different pressures.

The values of TDOS at Fermi level, $N(E_F)$ (states/eV-formula unit) is a very important electronic parameter that controls a large number of charge transport, superconducting state and magnetic properties of metals. Values of $N(E_F)$ decreases with increasing pressure. This is a consequence of the pressure induced shift in the Fermi level to higher energy. At 25 GPa, $N(E_F)$ is very small. Further increase in pressure may lead to an increase in the TDOS at the Fermi level. The repulsive Coulomb pseudopotential, $\mu^*$ can be determined from the TDOS at the Fermi level. The computed values of $\mu^*$ at different pressures are presented in Table 10. We have calculated $\mu^*$ using the following equation [77]:

$$\mu^* = \frac{0.26 N(E_F)}{1 + N(E_F)} \tag{14}$$

**Table 10.** TDOS at the Fermi level $N(E_F)$ (states/eV-formula unit) and repulsive Coulomb pseudopotential $\mu^*$ of $ScYH_6$ under different pressures.

| Compound | Pressure, P (GPa) | $N(E_F)$ | $\mu^*$ |
|---|---|---|---|
| $ScYH_6$ | 0 | 0.440 | 0.079 |
|  | 5 | 0.430 | 0.078 |
|  | 10 | 0.181 | 0.040 |
|  | 15 | 0.177 | 0.039 |
|  | 20 | 0.174 | 0.039 |
|  | 25 | 0.180 | 0.040 |

Typical values of $\mu^*$ lie within the range 0.10 to 0.20 [75-78]. Thus the Coulomb pseudopotential of $ScYH_6$ is quite low and decreases with increasing pressure. This also implies that the electronic correlations in this compound are rather weak.

*3.7 Superconducting state properties*

We have studied the superconducting critical temperatures of $ScYH_6$ under different uniform pressures using the widely applied McMillan equation [79] given below:

$$T_c = \frac{\theta_D}{1.45} \exp\left[-\frac{1.04(1+\lambda_{ep})}{\lambda_{ep} - \mu^*(1 + 0.62\lambda_{ep})}\right] \tag{15}$$

The computed values of the Debye temperatures and Coulomb pseudopotentials have been used to estimate the pressure dependent values of $T_C$. The electron-phonon coupling constants at different pressures have been taken from a previous report [1]. The calculated values of $T_C$ at different pressures are presented together with the relevant parameters in Table 11.

**Table 11**. The predicted superconducting transition temperatures and related parameters at different pressures of $ScYH_6$.

| Pressure, P (GPa) | $\lambda_{ep}$ [1] | $\mu^*$ | $\theta_D$ (K) | $T_C$ (K) | $T_C$ (K)$^\dagger$ [1] |
|---|---|---|---|---|---|
| 0  | 0.755  | 0.079 | 793.31 | 31.45 | 32.11 |
| 5  | 0.738♣ | 0.078 | 817.95 | 31.19 | 32.88♣ |
| 10 | 0.728♣ | 0.040 | 838.02 | 39.21 | 33.44♣ |
| 15 | 0.724♣ | 0.039 | 855.66 | 39.97 | 34.20♣ |
| 20 | 0.720♣ | 0.039 | 870.17 | 40.50 | 34.80♣ |
| 25 | 0.717♣ | 0.040 | 882.38 | 40.64 | 35.65♣ |

♣Interpolated from the values given in Ref. [1]
$^\dagger$Calculated with $\mu^* = 0.13$ [1]

From Table 11, we see that considering the possible errors in the interpolated values of the electron-phonon coupling constant, $\lambda_{ep}$, from Ref. [1], the agreement between the $T_C$ values calculated here and those in Ref. [1] is quite good. Our results suggest that the pressure induced increase in the superconducting transition temperature is due to the increase in the Debye temperature and decrease in the repulsive Coulomb pseudopotential in the pressure range considered. From Table 11 it also appears that $ScYH_6$ is a moderately coupled electron-phonon superconductor for pressures within 0 to 25 GPa. It should be noted that theoretical variation of $T_C$ for pressures up to 200 GPa with structural phase transition was considered in the previous work [1]. In this work, we focused our attention on much lower pressure behavior of $ScYH_6$, solely in the cubic (Pm-3) structure.

*3.8 Optical properties*

Optical properties of $ScYH_6$ are calculated using the electronic band structures. All the energy dependent optical constants are shown for 0 GPa and 25 GPa with the electric field polarization direction [100]. The dielectric function is a complex quantity given by:

$$\epsilon(\omega) = \epsilon_1(\omega) + i\epsilon_2(\omega) \tag{16}$$

Here, $\omega$ is the photon angular frequency, $\varepsilon_1$ and $\varepsilon_2$ are the real and imaginary parts of the dielectric function, respectively. The imaginary part of dielectric function can be expressed as [31,80]:

$$\varepsilon_2(\omega) = \frac{2\pi e^2}{\Omega \varepsilon_0} \sum_{k,v,c} |\psi_k^c| \boldsymbol{u}.\boldsymbol{r}|\psi_k^v|^2 \delta(E_k^c - E_k^v - E) \tag{17}$$

Here, $e$ is the electron charge, $\Omega$ is the unit cell volume, **u** is the unit vector along the polarization of the incident electric field, and $\psi_k^c$ and $\psi_k^v$ are wave functions for conduction and valence band electrons at a particular wave vector $k$, respectively. The real part of the dielectric constant is connected to the imaginary part and can be obtained via the Kramers-Kronig transformation equation. Once we have the complete complex dielectric function, all the other optical parameters can be extracted from it.

In metallic systems, both interband and intraband electronic transitions contribute to the imaginary part of the dielectric constant. Therefore, a Drude damping term of 0.05 eV has been included in its calculation to describe the low energy (far infrared) behavior [81-84]. Besides, a screened plasma energy of 3.0 eV has been used for the optical parameters computations.

Figure 8 (a) shows the real and imaginary parts of the dielectric function for photon energies up to 10 eV. The largest negative value of the real part ($\varepsilon_1$) is observed in the infrared region. The imaginary part rises in the low energy region indicating the Drude like behavior as seen in metals. The secondary peak in the imaginary part ($\varepsilon_2$) appears at ~2.8 eV. At this particular energy of the electromagnetic radiation, dielectric loss it peaked due to interband electronic transitions. At higher energies, both real and imaginary parts fall gradually. The effect of pressure on the dielectric function's spectra is small.

Reflectivity, $R(\omega)$, is connected to the dielectric function as follows [85-88]:

$$R(\omega) = \left| \frac{\sqrt{\varepsilon(\omega)} - 1}{\sqrt{\varepsilon(\omega)} + 1} \right|^2 \tag{18}$$

Fig. 8 (b) represents the reflectivity spectra at different pressures. The low energy reflectivity of ScYH$_6$ is very high. The reflectivity falls sharply at ~1.1 eV and then increases and levels of at around 2.1 eV. In the range 2 – 10 eV, reflectivity changes a little and remains below 40%. This shows that ScYH$_6$ is a weak reflector of visible and ultraviolet light. The effect of pressure is insignificant on the reflectivity spectra.

The refractive index consists of two terms: one is the real part of the refractive index (*n*) and the other one is the imaginary part of the refractive index, also known as the extinction coefficient (*k*). These two parts are calculated with the help of dielectric function as follows [85-88]:

$$n(w) = \sqrt{\frac{\sqrt{\varepsilon_1(\omega)^2 + \varepsilon_2(\omega)^2} + \varepsilon_1(\omega)}{2}} \qquad (19)$$

$$k(w) = \sqrt{\frac{\sqrt{\varepsilon_1(\omega)^2 + \varepsilon_2(\omega)^2} - \varepsilon_1(\omega)}{2}} \qquad (20)$$

The refractive index is shown in Fig. 8 (c). The real part of refractive index is associated with the phase velocity of light within the medium while the imaginary part indicates the amount of attenuation when the electromagnetic wave traverses through a medium. The low-energy refractive index is high in infrared region. The extinction coefficient falls to zero ~1.1 eV. This implies that ScYH$_6$ becomes transparent of incident electromagnetic wave at this particular energy. Once again, we find that pressure has minimum effect on the refractive index.

The absorption coefficient was calculated using the following equation [85-88]:

$$\alpha(w) = \sqrt{\omega \frac{\sqrt{\varepsilon_1(\omega)^2 + \varepsilon_2(\omega)^2} - \varepsilon_1(\omega)}{2}} \qquad (21)$$

The absorption coefficient indicates about the optimum solar energy conversion efficiency and how far light (specific energy) penetrate into the material before being absorbed. The absorption spectra for ScYH$_6$ are shown in Figs. 8 (d). The absorption coefficient starts rising from 0 eV which supports the metallic nature as found in the DOS calculations. Absorption coefficient falls sharply around 1.0 eV confirming the transparent behaviour of ScYH$_6$ at this energy. In the visible to ultraviolet regions, the absorption coefficient increases almost linearly with energy. This indicates that ScYH$_6$ is a potentially efficient absorber of ultraviolet radiation.

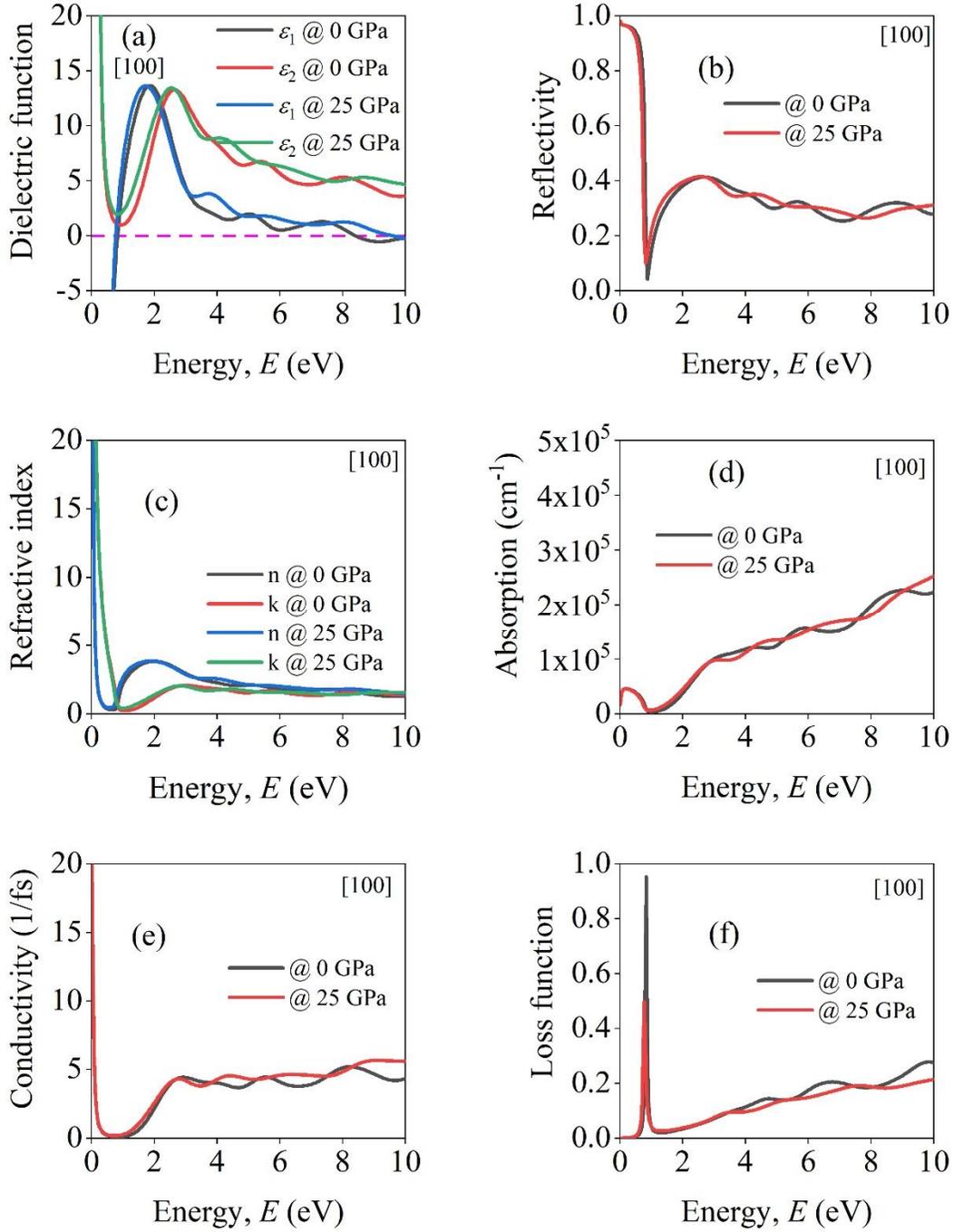

**Figure 8**: Optical properties of ScYH$_6$ at 0 GPa and 25 GPa with electric field polarization along [100] direction.

The optical conductivity, $\sigma(\omega)$, can also be calculated from the complex dielectric function using the following relation [85-88]:

$$\sigma(\omega) = -\frac{i\omega}{4\pi}\varepsilon(\omega) \qquad (22)$$

The real part of the optical conductivity of $ScYH_6$ is presented in Fig. 8 (e). The photoconductivity starts from zero photon energy which complements the result of metallic nature. Optical conductivity of $ScYH_6$ is quite high at low energy due to free electron contribution. The optical conductivity falls sharply at ~1.0 eV. Above 2.0 eV, the optical conductivity becomes almost nonselective.

The energy loss spectra are presented in Fig. 8 (f). This parameter can also be obtained from the dielectric constant [85-88]. The loss function is an important optical parameter which is used to understand the screened optical excitation spectra. The peak in the energy loss spectrum appears at a particular incident light energy that gives the information about the bulk plasma frequency The plasma peak is found at ~1.0 eV for $ScYH_6$. The plasma energy is quite low; this is a consequence of relatively low charge carrier concentration of this compound. At the plasma energy, the optical properties of $ScYH_6$ become similar to those of an insulator. Like in all other optical parameters, the effect of pressure on the loss function is also insignificant.

## 4. Conclusions

In this study we have explored the structural, electronic, elastic, mechanical, thermophysical, superconducting and optical properties of cubic $ScYH_6$ (Pm-3 structure) under pressure. Most of the results presented herein are novel. The structural parameters are in good agreement with the previous work [1]. It has been found that $ScYH_6$ is thermodynamically stable. $ScYH_6$ is brittle in nature; the brittleness decreases with pressure. Electronic band structure exhibits weak metallic nature. The compound has high electronic and structural stability. The TDOS at the Fermi level is low and so are the electronic correlations. The elastic anisotropy is low. The compound under study is fairly hard and moderately machinable. The Debye temperature of $ScYH_6$ is high. The superconducting $T_C$ shows positive correlations with pressure as found in a previous study [1]. Optical parameters are studied in detail. The effect of pressure on the optical parameters' spectra is weak. The compound is an efficient absorber of ultraviolet light. It is also a very efficient reflector of infrared light. The optical spectra reflect the underlying electronic band structure features very well.

**Acknowledgements**

S. H. N. acknowledges the research grant (1151/5/52/RU/Science-07/19-20) from the Faculty of Science, University of Rajshahi, Bangladesh, which partly supported this work. Md. A. A. acknowledges the financial support from the Bangabandhu Science and Technology Fellowship Trust for his Ph.D. research.

**Data availability**

The data sets generated and/or analyzed in this study are available from the corresponding author on reasonable request.

**Declaration of interest**

The authors declare that they have no known competing financial interests or personal relationships that could have appeared to influence the work reported in this paper.

# CRediT author statement

**Md. Ashraful Alam:** Methodology, Software, Formal analysis, Writing-Original draft. **F. Parvin:** Supervision, Writing-Reviewing and Editing. **S. H. Naqib:** Conceptualization, Supervision, Formal analysis, Writing- Reviewing and Editing.